\documentclass[12pt]{article}
\usepackage{epsfig}\parskip 5pt plus 1pt
\usepackage{amssymb}
\newcommand{\be}{\begin{equation}}
\newcommand{\ee}{\end{equation}}
\newcommand{\bea}{\begin{eqnarray}}
\newcommand{\eea}{\end{eqnarray}}

\def\simge{\mathrel{%
   \rlap{\raise 0.511ex \hbox{$>$}}{\lower 0.511ex \hbox{$\sim$}}}}
\def\simle{\mathrel{
   \rlap{\raise 0.511ex \hbox{$<$}}{\lower 0.511ex \hbox{$\sim$}}}}

\newcommand{ \mysmall}[1]{\scriptscriptstyle #1} 

\begin{document}
\thispagestyle{empty}
\vspace*{1cm}
\begin{center}
{\Large{\bf Testing the $a_\mu$ anomaly in the electron sector through 
a precise measurement of $h/M$ } }\\

\vspace{.5cm}
F.~Terranova$^{\rm a}$, G.~M.~Tino$^{\rm b}$ \\
\vspace*{1cm}
$^{\rm a}$ Dep. of Physics, Univ. of Milano-Bicocca and INFN, Sezione di Milano-Bicocca, 
Milano, Italy \\
$^{\rm b}$ Dep. of Physics and Astronomy and LENS Laboratory, Univ. of Florence,
INFN Sezione di Firenze, Sesto Fiorentino (FI), Italy \\
\end{center}

\vspace{.3cm}
\begin{abstract}
\noindent
The persistent $a_\mu \equiv (g-2)/2$ anomaly in the muon sector could
be due to new physics visible in the electron sector through a sub-ppb
measurement of the anomalous magnetic moment of the electron
$a_e$. Driven by recent results on the electron mass (S.~Sturm {\it et
  al.}, Nature 506 (2014) 467), we reconsider the sources of
uncertainties that limit our knowledge of $a_e$ including current
advances in atom interferometry. We demonstrate that it is possible to
attain the level of precision needed to test $a_\mu$ in the naive
scaling hypothesis on a timescale similar to next generation $g-2$
muon experiments at Fermilab and JPARC. In order to achieve such level
of precision, the knowledge of the quotient $h/M$, i.e. the ratio
between the Planck constant and the mass of the atom employed in the
interferometer, will play a crucial role. We identify the most
favorable isotopes to achieve an overall relative precision below
$10^{-10}$.
\end{abstract}

\vspace*{\stretch{2}}
\begin{flushleft}
  \vskip 2cm
{ PACS: 6.20.Jr, 13.40.Em, 03.75.Dg  \\ } 
\end{flushleft}

\newpage

\section{Introduction}
\label{introduction}

In the last 40 years, the experimental accuracy on the anomalous
magnetic moment of the muon $a_\mu= (g-2)_\mu /2$ has been improved by
more than five orders of magnitudes~\cite{Jegerlehner:2009ry}.  The final
results of the Fermilab E821 experiment~\cite{bnl} shows a clear
discrepancy with respect to the Standard Model (SM) prediction,
corresponding to a $\sim$3.5~$\sigma$ deviation. Such a puzzling outcome
has boosted a vigorous experimental programme and new results from the
E989 Fermilab~\cite{E989} and g-2 JPARC~\cite{JPARC} experiments are
expected in a few years. If the origin of the muon discrepancy is due
to physics beyond SM, similar effects are expected in the electron
sector, too. In particular, corrections due to new physics should
appear in the electron magnetic moment $a_e= (g-2)_e /2$. In general,
these corrections will be suppressed by a $\mathcal{O} [ (m_e/m_\mu)^2 ] $
factor with respect to muons (``naive scaling''), being $m_e$ and
$m_\mu$ the mass of the electron and muon, respectively.

In fact, progresses in the measurement and theoretical understanding
of $a_e$ are so impressive that $a_e$ could be implemented as an
observable to test new physics in the electroweak sector of the
Standard Model~\cite{Giudice:2012ms}. Driven by recent results on the
electron mass~\cite{sturm_nature}, in this paper we reconsider the
sources of uncertainties that limit our knowledge of $a_e$ including
current advances in atom interferometry. We demonstrate that it is
possible to attain the level of precision needed to test $a_\mu$ in
the naive scaling hypothesis on a timescale similar to next generation
$g-2$ muon experiments and we identify the best experimental strategy
to reach this goal.

In particular, in Sec.~\ref{sec:NP} we discuss the electron
counterpart of new physics effects that can generate the muon
discrepancy and we set the scale for the experimental precision that
has to be attained. The observables that must be determined with high
precision are discussed in Sec.~\ref{sec:observables}: they are
$a_e^{exp}$ (Sec.~\ref{sec:ae_exp}), the fine structure constant
$\alpha$ (Sec.~\ref{sec:alpha}) and four ancillary observables: the
Rydberg constant $R_\infty$ (Sec.~\ref{sec:alpha}), the electron mass
in atomic mass units (Sec.~\ref{sec:ele_mass}), the mass of the atom
employed in the atomic interferometer (Sec.~\ref{sec:atom_mass}) and
the ratio between the Planck constant and the atom mass ($h/M$,
Sec.~\ref{sec:interferometry}). For each of these observables we
determine the best current accuracy and the improvements that are needed
to reach the goal sensitivity. The sensitivity to new physics of $a_e$
and the comparison with new physics effects in the muon sector are
discussed in (Sec.~\ref{sec:sensitivity}).

\section{The $a_\mu$ anomaly and its electron counterpart}
\label{sec:NP}

The precise measurement of the anomalous magnetic moment of the
electron $a_e= (g-2)_e /2$ is one of the most brilliant test of QED
and a key metrological observable in fundamental physics. In the last
twenty years the relative experimental precision on $a_e$ reached 
sub-ppb precisions (see Sec.\ref{sec:ae_exp}). Progress
in theory predictions matches the improved precision of the
measurements and, given an independent determination of the fine
structure constant $\alpha$, $a_e$ provides a clean test of
perturbative QED at five-loop level. Conversely, if we assume QED to
be valid, $a_e$ offers the most precise measurement of $\alpha$
available to date and drives the overall CODATA fits both for $\alpha$
and for several correlated quantities as the molar Planck
constant~\cite{Mohr:2012tt}.  In high-energy-physics, $a_e$ plays a
role both as a constraint for $\alpha(q^2 \rightarrow
0)$~\cite{ref:PDG} and for the determination of the QED contributions
to the muon anomaly $a_\mu$.  In fact, progresses in the measurement
and theoretical understanding of $a_e$ are so impressive that $a_e$
could be implemented as an observable to test new physics in the
electroweak sector of the Standard Model~\cite{Giudice:2012ms}. Up to
now, this role pertains solely to $a_\mu$ since new physics (NP)
effects in $a_\mu$ and $a_e$ (loop effects due to a new physics scale
$\Lambda_\mu$ and $\Lambda_e$) decouple as $(m_\mu/\Lambda_\mu)^2$ and
$(m_e/\Lambda_e)^2$, respectively. The case where $\Lambda_{\mu}\equiv \Lambda_{e}$ is referred to as
``naive scaling'' ({\small NS}) and, when NS is at work,
we thus expect $a_\mu$ to be $(m_\mu/m_e)^2$ more sensitive to NP than its electron counterpart. On
the other hand, $a_e$ is currently measured $\sim$2300 times more accurately
than $a_\mu$ and further improvements are expected in the years to
come.  These considerations have led the authors of
Ref.~\cite{Giudice:2012ms} to evaluate the physics potential of $a_e$
as a probe of new physics both in the naive scaling approximation and
in specific models where naive scaling is violated. 

The main motivation to promote $a_e$ to a probe for new physics is the
above-mentioned discrepancy in the measurement of $a_\mu$.  The final
results of the Fermilab E821 experiment sets the scale of the $a_\mu$
discrepancy. It amounts to a shift with respect to the Standard Model (SM)
prediction of~\cite{bnl}
\be
\Delta a_\mu = a_\mu^{\mysmall \rm EXP}-a_\mu^{\mysmall \rm SM} = 2.90 \, (90) \times 10^{-9}\,,
\label{eq:gmu_exp}
\ee
corresponding to a $3.5~\sigma$ discrepancy. If the discrepancy is due to
new physics, we can always parametrize such NP effects as follows,
\be
|\Delta a_\mu| = \frac{m^2_\mu}{\Lambda^{2}_{\mu}}\,,
\label{eq:gmu_NP}
\ee
where the NP scale $\Lambda_{\mu}$ encodes possible loop factors, loop functions
and couplings of new particles to the muon. As a result, the central value of 
Eq.~(\ref{eq:gmu_exp}) can be accommodated for $\Lambda_{\mu}\approx 2~$TeV.
Defining the NP effects for the electron $g-2$ analogously to Eq.~\ref{eq:gmu_NP},
it turns out that
\be
\left|\frac{\Delta a_e}{\Delta a_\mu} \right|=
\frac{m^2_e}{m^2_\mu}\frac{\Lambda^{2}_{\mu}}{\Lambda^{2}_{e}}\,.
\label{eq:gm2_ratios}
\ee
Assuming {\small NS},
the $a_\mu$ discrepancy could be tested in the electron sector once the
experiments reach a precision of
\be
\sigma_{a_e} = 2.9 \times 10^{-9} \times \left(\frac{m_e}{m_\mu}\right)^2 =
6.8 \times 10^{-14}~ (0.06~{\rm ppb})\,.
\label{eq:ge_exp}
\ee However, in concrete examples of new physics theories, NS could be
violated and larger effects in $a_e$ might be expected. For instance,
in supersymmetric theories~\cite{Giudice:2012ms,Aboubrahim:2014hya}
with non degenerate slepton masses $m_{\tilde e} \neq m_{\tilde \mu}$,
we can identify $\Lambda_{\mu} \equiv m_{\tilde \mu}$ and $\Lambda_{e}
\equiv m_{\tilde e}$ and $\Delta a_e$ can even saturate the current
experimental bound $\Delta a_e \approx 10^{-12}$.

In spite of the progresses in the experimental determination of $a_e$,
it is generally believed that new physics effects that could be
responsible for the $a_\mu$ anomaly will be observed in the electron
sector only in the occurrence of a strong violation of naive scaling
($\Lambda_\mu \gg \Lambda_e$).  This is due to the fact that $a_e$ is
deeply entangled with most of the fundamental constants in physics and
a direct observation of NP in the lepton sector requires an
independent determination of such constants. As discussed in the
following, latest advances in metrology remove most of these obstacles
and make possible to attain a level of precision close to the target
of Eq.~\ref{eq:ge_exp}.

\section{Experimental observables}
\label{sec:observables}

\subsection{The experimental determination of $a_e$}
\label{sec:ae_exp}

In the last twenty years the experimental precision achieved on $a_e$
by cylindrical Penning traps has improved by more than one order of
magnitude the one of hyperbolic traps and the opportunities offered by
these techniques have not been fully exploited,
yet~\cite{ref:gabrielse_book}.

The best world measurement of $a_e$, i.e. the 2010 Harvard measurement
with a cylindrical Penning trap~\cite{Hanneke:2010au}, achieved a
relative accuracy of 0.24~ppb. This uncertainty is four times
larger than the precision needed to observe NS effects in the electron
sector.  Still, cylindrical Penning traps have not saturated their
systematics and major improvements can be envisaged. The 2006 Harvard
measurement~\cite{Odom:2006zz} was mostly dominated by cavity shift
modeling. In cylindrical Penning traps the interaction of the trapped
electron and the cavity modes shifts the cyclotron frequency and the
shift must be properly modeled to extract $a_e$. This drawback is
unavoidable if spontaneous emission of radiation has to be inhibited.
Note, however, that the current measurement of
$a_e$~\cite{Hanneke:2008tm} is not dominated by the cavity shift, yet;
systematics arise from an anomalous broadening of the spectroscopy
lineshapes (probably due to fluctuations in the magnetic
field~\cite{Hanneke:2010au}) and to statistics.  In particular,
lineshape modeling accounts for most of the systematic budget of
$a_e$. A breakdown of the systematics of the 2008 analysis is
available in Tab.~6.6 of Ref.~\cite{hanneke_thesis}, where the
(run-to-run correlated) lineshape model uncertainty accounts for a
relative uncertainty of 0.21~ppb on $a_e$. The overall uncertainty on
$a_e$ reported in Ref.~\cite{Hanneke:2010au} is 0.24~ppb (0.28~ppt in
$g/2$).

Most likely, experiments based on cylindrical Penning traps will be
ultimately limited by cavity shift uncertainties, which are a source
of systematics intrinsic to this technology.  Results from the Harvard
Group~\cite{Hanneke:2010au,hanneke_thesis} indicate that this
contribution to the relative uncertainty on $a_e$ can be reduced below
$0.08$~ppb ($<0.1$~ppt for g/2).

\subsection{The fine structure constant and the link to $h/M$}
\label{sec:alpha}

The Harvard measurement of $a_e^{exp}$ would already be able to
constrain specific models that break NS and enhance new physics
contributions in $a_e$ with respect to $a_\mu$. However, the
measurement becomes quite marginal once we diagonalize the correlation
matrix between $a_e^{exp}$ (which is commonly used to extract
$\alpha$) and its theory expectation $a_e^{SM}$. This is due to the
fact that $a_e^{SM}$ is $\alpha/2\pi$ at leading order and, hence, it
is highly dependent on $\alpha$.  If we resort to a fully independent,
albeit less precise, determination of $\alpha$, the accuracy on the
theory prediction for $a_e$ ($a_e^{SM}$) is worsened to
0.66~ppb~\cite{Giudice:2012ms}.

The possibility of having a ppb measurement of $\alpha$ independent of
$a_e$ became viable
with the measurement of the narrow (1.3~Hz) 1S-2S two-photon resonant
line of hydrogen atom with a relative precision of $3.4 \times
10^{-13}$~\cite{Udem:1997zz} and with the precise measurement of the
$h/M$ ratio by atom interferometry~\cite{Cronin2009,Tino2013b}. 
The new data on hydrogen spectroscopy resulted in a measurement of the
Rydberg constant with a precision better than 0.01~ppb ($7 \times
10^{-12}$~\cite{Mohr:2012tt}). Since $R_{\infty} = m_e \alpha^2 c/2h$,
the outstanding precision on $R_{\infty}$ links $\alpha$ to the
evaluation of the quotient $h/m_e$.  In fact, for a given atom $X$
whose mass is $M_X$,
\be 
\alpha^2 \ = \ \frac{2R_{\infty}}{c} \frac{M_X}{m_e}\frac{h}{M_X}
 \ \ = 2 \frac{R_{\infty}}{c} \frac{M_X}{m_u} \frac{m_u}{m_e}
 \frac{h}{M_X} \ .
\label{eq:rydberg}
\ee
$m_u$ being the atomic mass units, i.e.  the mass of 1/12 of the mass
of $^{12}$C.  Eq.~\ref{eq:rydberg} paved the way for an independent
determination of $\alpha$ based on cold atom interferometry (see
Sec.~\ref{sec:interferometry}) since atom interferometers are well
suited to measure the quotient $h/M_X$. This quotient is also
of great metrological interest for the redefinition of the
kilogram~\cite{bouchendira_review}.  On the other hand, the
exploitation of the Rydberg relationship to extract $\alpha$ causes
two additional sources of uncertainties: the systematics due to the
knowledge of the mass ($M_X/m_u$) of the atom employed to measure
$h/M_X$ and the uncertainty on $m_e$ in atomic mass unit ($m_e/m_u$).

\subsection{The electron mass}
\label{sec:ele_mass}

The most straightforward way to employ Eq.~\ref{eq:rydberg} with
minimum penalty from the knowledge of masses would be to design an
ancillary experiment aimed at determining the mass ratio between the
electron and the isotope employed in the atomic interferometer. The
expected uncertainty for such dedicated experiment would be of the
order of the {\it direct} measurement of $A_r(m_e) \equiv
m_e/m_u$~\cite{Mohr:2012tt}.  In fact, while the most precise
evaluation of $A_r(m_e)$ can be obtained from bound-state
electrons~\cite{sturm_nature} (see below), the best direct measurement
of $A_r(m_e)$ remains the one of the Washington
Univ. experiment~\cite{Farnham:1995zz}. Here, the cyclotron frequency
of an electron and a $^{12}\mathrm{C}^{6+}$ ion were compared in a
Penning trap. The measurement determined $A_r(m_e)$ with just a
2.1~ppb relative accuracy and, at present, this technique does not seem
appropriate to reach the target of Eq.~\ref{eq:ge_exp}.

Since the 2002 CODATA adjustment, $A_r(m_e)$ is determined from the
measurement of $a_e$ in bound-state QED. For a bound electron in
hydrogen-like systems of nuclear charge Z, the electron anomalous
magnetic moment $g_b$ is perturbed with respect to the free-particle
value. The leading order (pure Dirac) contribution is $g=2$ for a free
electron and $g_b^{Breit}=2/3 ( 1+
2\sqrt{1-Z^2\alpha^2})$~\cite{breit} for a bound electron in a field
generated by an atom with nuclear charge $Z$. In the past, the best
measurements of $g_b$ for $\mathrm{C}^{5+}$ and $\mathrm{O}^{7+}$ ions have been
performed at GSI using a Penning trap to measure the ratio between the
Larmor $\omega_L$ and cyclotron $\omega_c$ frequencies of the stored
ions~\cite{werth2006}. The relationship between $g_b$ and these
frequencies
\be
g_b=2\frac{\omega_L}{\omega_c} \frac{m_e}{M_{C^{5+}/O^{7+}}} \ = \ 
2\frac{\omega_L}{\omega_c} A_r(m_e) \frac{m_u}{M_{C^{5+}/O^{7+}}} 
\label{eq:g_b}
\ee
links $g_b$ with the electron mass. Very recently~\cite{sturm_nature},
the GSI group has improved by a factor of 13 the CODATA value
performing a measurement of the frequency ratio for the hydrogen-like
atom $^{12}\mathrm{C}^{5+}$. The corresponding estimate for $m_e$ relies on
bound-QED calculations~\cite{bound_qed_theory}, which can be checked
by ancillary $g_b$ measurements on
$^{28}\mathrm{Si}^{13+}$~\cite{sturm2011,sturm2013}.

The GSI evaluation of $m_e$ in atomic mass units is:
\begin{equation}
\frac{m_e}{m_u} = 0.000548579909067(14)(9)(2) 
\end{equation}
where the errors in parentheses are the statistical error, the
experimental systematic uncertainty and the theoretical (bound-QED)
error, respectively.  This measurement provides $m_e/m_u$ with a
relative precision of 0.03~ppb, which scales to an uncertainty on
$\alpha$ of 0.015~ppb. This error is well below the systematic budget
to observe the NS muon anomaly in the electron sector.

\subsection{$M/m_u$}
\label{sec:atom_mass}

The only remaining source of uncertainty due to the Rydberg
relationship (\ref{eq:rydberg}) is the knowledge of the $M_X/m_u$
ratio.  It is therefore interesting to estimate current uncertainties
on $M_X/m_u$ for viable cold atom candidates and evaluate whether the $M_X/m_u$
error can be improved with an appropriate choice of the isotope $X$.
For what concerns alkali-metal atoms, the atomic masses of the
isotopes relevant for the determination of $\alpha$ have been measured
with high precision employing orthogonally compensated Penning
traps. Recent results~\cite{Myers2013,mount2010} from Washington
Univ. determine the atomic masses of
$^{23}\mathrm{Na}$,$^{39,41}\mathrm{K}$, $^{85,87}\mathrm{Rb}$ and
$^{133}\mathrm{Cs}$ with a precision of $\simeq$0.1~ppb.  In
particular, the error associated with the current best atom for
the measurement of $h/M$ ($^{87}\mathrm{Rb}$) is
0.115~ppb~\cite{mount2010}. It corresponds to a relative
uncertainty on $\alpha$ of 0.06~ppb. Hence, performing experiments
with isotopes different from $^{87}\mathrm{Rb}$ but chosen among the
standard alkali metal candidates do not bring to sizable improvements
on $\alpha$.  A notable exception is $^4\mathrm{He}$, which is simultaneously a
good candidate for atom interferometry (see
Sec.~\ref{sec:interferometry}) and whose mass is known with
outstanding accuracy (0.015~ppb \cite{VanDyck:2004zz}). The use of
$^4$He would allow the exploitation of the Rydberg $\alpha$-to-$h/M$
link without any significant penalty since the overall relative error
due to $m_e$ and $M_{He}$ impacts on $\alpha$ at the
$\frac{1}{2}(3  \oplus 1.5)  \times 10^{-11} = 0.017$~ppb
level.

\subsection{The $h/M$ quotient}
\label{sec:interferometry}

In order to test the muon anomaly in the electron sector, the
measurement of the quotient $h/M_{X}$ remains the main source of
uncertainty and a remarkable experimental challenge. The most precise
value obtained so far by atom interferometers is for
$^{87}\mathrm{Rb}$~\cite{Bouchendira2011} and it corresponds to
$h/M_{Rb} = 4.5913592729(57) \times 10^{-9}$~m$^2$s$^{-1}$
(1.24~ppb). The error budget of \cite{Bouchendira2011} updated with
the latest determination of $m_e/m_u$~\cite{sturm_nature} and
$M_{Rb}/m_u$~\cite{mount2010} is thus
\be 
\frac{\sigma_\alpha}{\alpha} \simeq \frac{1}{2} \left[
  1.24 \oplus 0.03 \oplus 0.115
\right] \ \mathrm{ppb} = 0.62~\mathrm{ppb} \, 
\ee
well above the scale needed to test the $a_\mu$ discrepancy in the NS
framework.  However, there is plenty of scope for improvement in the
measurement of $h/M_X$ and the potential of the experimental technique
is still to be fully exploited.

The very first measurement of $\alpha$ employing atom interferometry
was carried out in Stanford~\cite{Wicht2002} using Cesium atoms. The value
of $\alpha$ was measured with a relative precision
$\sigma_\alpha/\alpha = 7.4 \times 10^{-9}$ mainly limited by possible
index of refraction effects in the cold atoms sample.  More recently,
in an experiment at Berkeley, Cs is used in an interferometer based on
a Ramsey-Bord{\'e} scheme.  At present, the achieved relative
uncertainty for $\alpha$ is $\sim$2~ppb~\cite{Lan2013}. The error here
is mostly statistical (1.7 ppb); the next largest error is a parasitic
phase shift caused by the beam splitters in the simultaneous conjugate
interferometers. Recently, using Bloch oscillations to increase the
separation of the interferometers, the signal to noise ratio was
improved by about one order of magnitude and the parasitic phase shift
reduced, so that a precision below ppb should be within reach
\cite{Mueller}.

As already mentioned, the most accurate determination of $h/M_X$ has
been obtained with Rubidium atoms. Several experiments were performed
\cite{Bouchendira2011,Clade2006,Cadoret2008} in Paris with Rb
using an atom interferometer based on a combination of a
Ramsey-Bord{\'e} interferometer \cite{Borde1989} with Bloch
oscillations. The precision in \cite{Bouchendira2011} is mostly
limited by laser beams alignment, wave front curvature and Gouy phase
effects.  A new project is ongoing, which is aimed to improve the
accuracy on $h/M_{Rb}$ and therefore on $\alpha$ by increasing the
sensitivity of the atom interferometer and reducing the systematic
effect due to the Gouy phase and the wavefront
curvature~\cite{Clade}. Key elements of the new experiment will be the
use of evaporatively cooled atoms and an atom interferometer based on
large momentum beam splitters \cite{Bouchendira2013}.

Future prospects are the use of other atoms such as Helium or
Strontium. An experiment on He was started in Amsterdam
\cite{Vassen}.  It is based on metastable $^{4}\mathrm{He}$ in a
1D-lattice setup to perform Bloch oscillations and velocity
measurement with an atom interferometer. Metastable $^{4}\mathrm{He}$
has some advantages compared to Rb and other atoms. These relate to
the low mass, the smaller sensitivity to magnetic fields and the
availability of high-power infrared fiber lasers at the relevant
wavelength of 1083 nm. The use of a metastable state enables an
alternative detection on a microchannel plate detector but also causes
Penning ionization losses at high densities. Therefore helium has the
potential to become at least as accurate as Rb using the same method
and, as noted above, the helium mass is known with a relative
uncertainty of 0.015~ppb~\cite{VanDyck:2004zz,Myers2013}. The
potential of Sr for high precision atom interferometry was
demonstrated in experiments based on Bloch oscillations
\cite{Ferrari2006b,Poli2011}.  Because of the specific characteristics
of this atom, experiments with Sr using atom interferometry schemes as
the ones already demonstrated for different atoms promise to reach a
very high precision \cite{TinoinVarenna}.  The mass ratio for Sr
isotopes was measured in \cite{Rana2012} with a relative precision of
0.11~ppb.

These experimental programmes are aimed at  a precision in the
determination of $h/M_X$ $<$0.1~ppb.   On a much longer
timescale, a final precision of $10^{-11}$ could be achieved in a
space experiment where microgravity would allow to fully exploit the
potential sensitivity of atom interferometers \cite{Tino2013}.
  
\section{Sensitivity to new physics in the electron sector}
\label{sec:sensitivity}

\begin{figure}
\centering\includegraphics[width=0.8\textwidth]{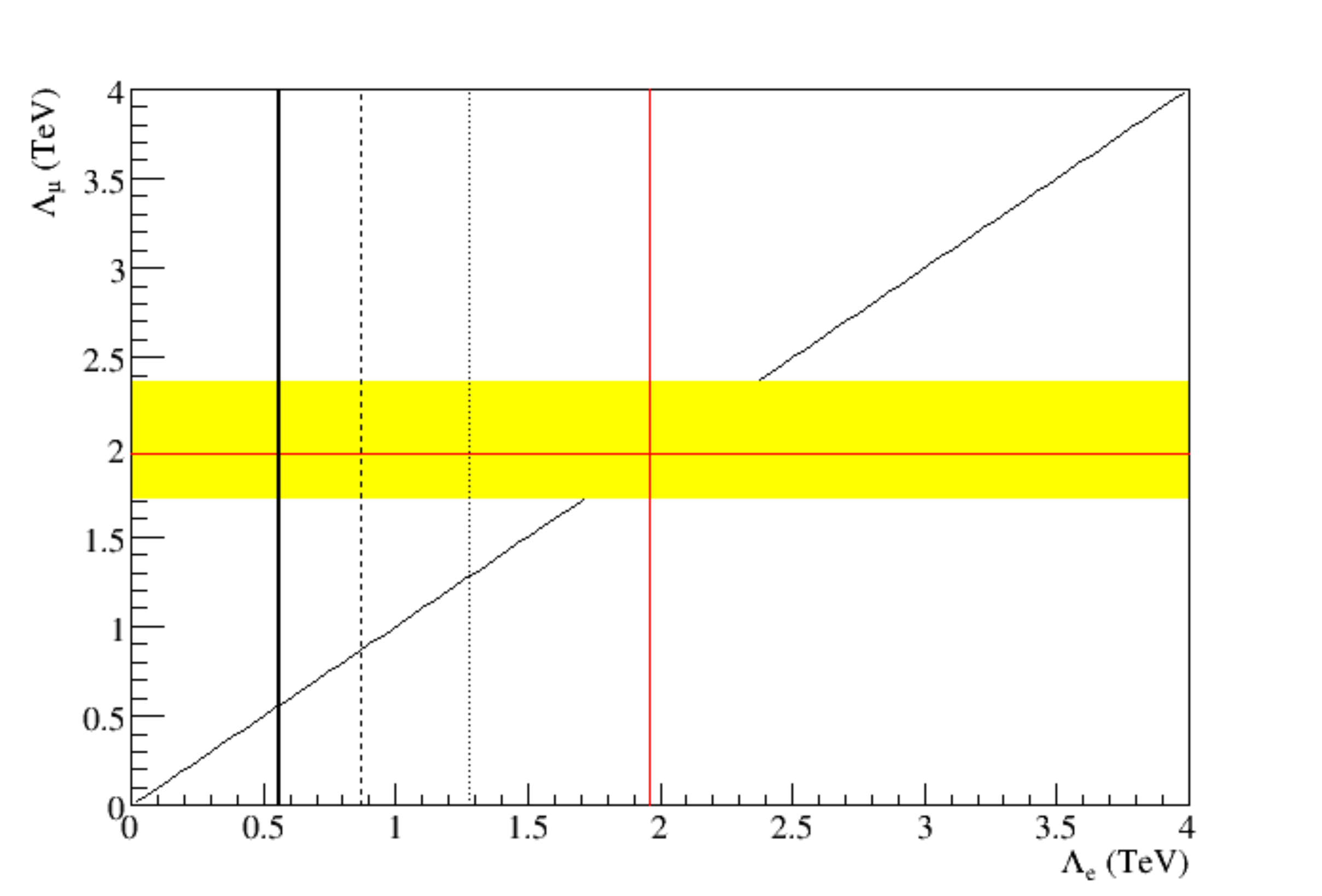} 
\caption{$\Lambda_\mu$ versus $\Lambda_e$ in TeV. The red horizontal
  line indicates the best fit of $\Lambda_\mu$ from the muon
  anomaly. The horizontal band is the corresponding 1~$\sigma$
  uncertainty. The area on the right of the thick vertical line shows
  the allowed values of $\Lambda_e$ from the the current accuracy on
  $a_e$ and assuming the NS expectation from the muon anomaly (red
  thin vertical line) as central value. The diagonal line corresponds
  to the NS expectation $\Lambda_\mu = \Lambda_e $. The tighter
  constraints on $\Lambda_e $ are computed removing the systematics
  from $\alpha$ (dashed line) and envisaging a reduction of the
  experimental uncertainty to the cavity shift systematics for
  Penning traps (dotted line).}
\label{fig:lambda}
\end{figure}

The relevance of the contributions discussed in previous sections can
be expressed in terms of constraints to the new physics scale
$\Lambda_e$, as depicted in Fig.~\ref{fig:lambda}.  The red horizontal
line indicates the best fit of $\Lambda_\mu$ from the muon anomaly:
$\Lambda_\mu = \sqrt{m_\mu^2/2.90 \times 10^{-9}} \simeq 2$~TeV.  The
horizontal band is the corresponding 1~$\sigma$ uncertainty. The three
vertical lines represent the constrains from the electron sector
computed under different assumptions on the systematics.  The total
uncertainty on $a_e$ (leftmost thick line) is computed using the 2010
Harvard measurement~\cite{Hanneke:2010au} and taking $\alpha$ from the
best measurement of the $h/M$ ratio
($h/M_{Rb}$~\cite{Bouchendira2011}). Assuming the NS expectation from
the muon anomaly as central value (thin red vertical line of
Fig.~\ref{fig:lambda}), such accuracy sets a limit of $\Lambda_e
\gtrsim 0.6$~TeV (thick black line of Fig.~\ref{fig:lambda}). In the
occurrence of NS, $\Lambda_e = \Lambda_\mu$ holds (diagonal line) and
a deviation of $a_e$ from its SM prediction is expected for $|\Delta
a_e| \simeq \sigma_{a_e} \simeq 6.8 \times 10^{-14}$.  The tighter
constraints on $\Lambda_e $ are computed removing the systematics from
$\alpha$ (dashed vertical line) and envisaging a reduction of the
experimental uncertainty to the cavity shift systematics for Penning
traps (dotted vertical line).

If $\alpha$ can be disentangled from $a_e^{exp}$ at the appropriate
level of precision, perspectives to test new physics in the $a_e$
sector are very encouraging. In fact, the uncertainty in the
theoretical determination of $a_e^{SM}$ is appropriate to test the
$a_\mu$ anomaly at NS level. Such uncertainty mostly resides in the
numerical approximation employed to evaluate four and five
loop QED contributions (0.06~ppb)~\cite{Aoyama:2012wj} and in the
hadronic term (0.02~ppb)~\cite{Nomura:2012sb}. The overall amount is
within the error budget for NS (0.06~ppb) and further improvements are
in reach.

\begin{figure}
\centering\includegraphics[width=0.8\textwidth]{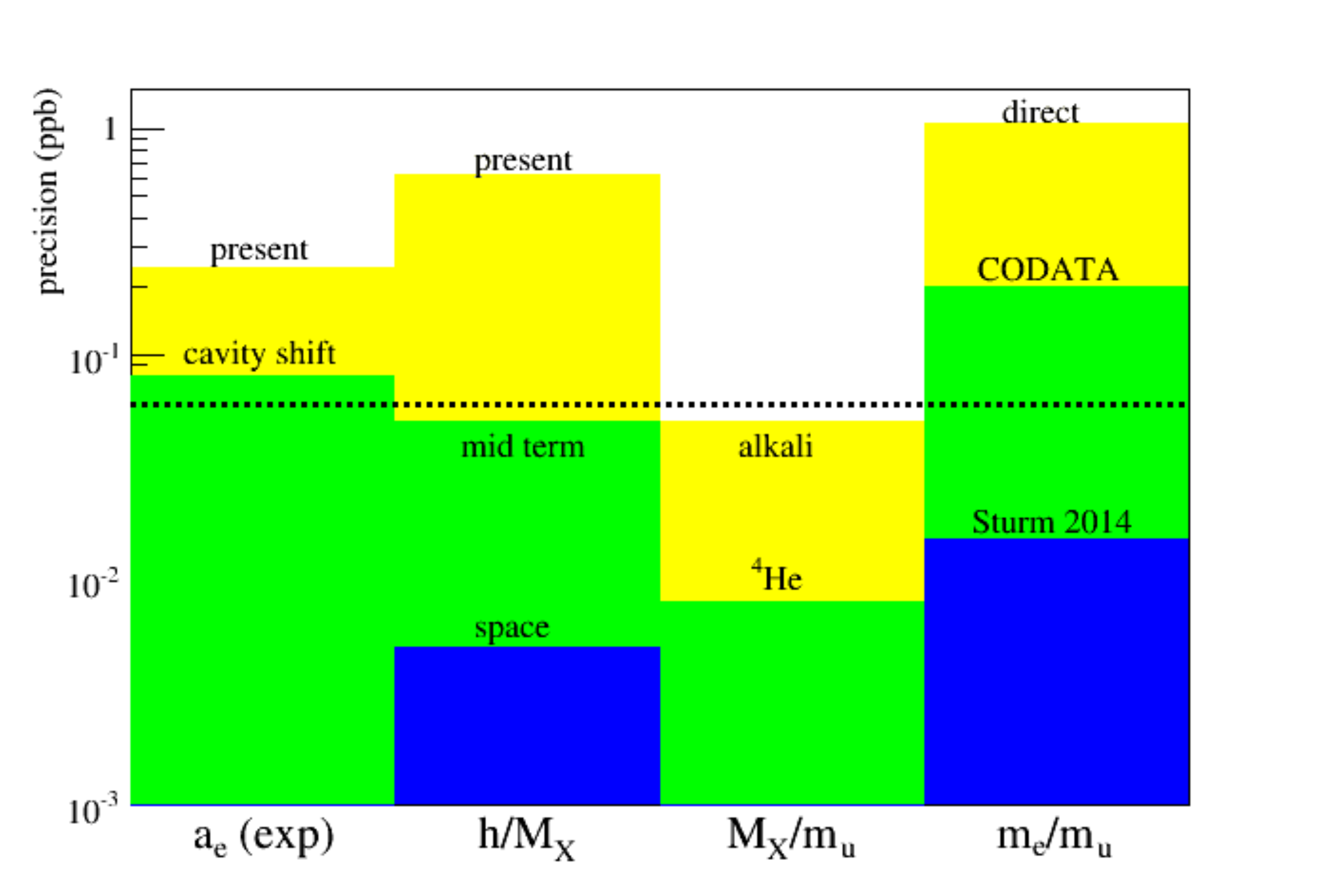} 
\caption{Summary of the contributions to the relative precision on
  $a_e$ (in ppb). $a_e$ (exp) is the experimental contribution from
  the measurement of $a_e$ with Penning traps. $h/M_X$ is the
  contribution from the quotient $h/M$ measured with atom
  interferometers for an isotope $X$. The contribution due to the
  knowledge of the isotope mass in atomic mass units is labeled
  $M_X/m_u$. The uncertainty on $A_r(m_e) \equiv m_e/m_u$ is shown in
  the $m_e/m_u$ column. Here, ``direct'' refers to the direct
  measurement of~\cite{Farnham:1995zz}; the CODATA 2010 value is
  labeled ``CODATA'' and the recent GSI
  measurement~\cite{sturm_nature} is labeled ``Sturm 2014''. The
  horizontal line corresponds to the NS size of the expected anomaly
  (0.06~ppb).}
\label{fig:summary}
\end{figure}

The experimental systematics budget is summarized in
Fig.~\ref{fig:summary}.  As mentioned above, the technology of the
cylindrical Penning trap (first column of Fig.~\ref{fig:summary}) can
be pushed below the current cavity shift limit (0.08~ppb) to reach the
NS precision range (0.06~ppb - horizontal line of
Fig.~\ref{fig:summary}). However, such measurement can be effective
only if an independent measurement of $\alpha$ is available with a
precision $<$0.1~ppb. The outstanding accuracy reached on the Rydberg
constant allows to obtain such measurement from atom interferometers
through a precision measurement of the $h/M$ quotient (second
column). This experimental approach highly profits from recent
advances in the measurement of the electron mass, which was considered
as a possible limiting factor in the past (fourth column).  Atom
interferometers based on alkali atoms are able to reach the requested
accuracy although they introduce an additional source of uncertainty
due to the error on $M_X/m_u$ (third column). This systematics is
negligible in $^4\mathrm{He}$-based interferometers.

\section{Conclusions}
\label{sec:conclusions}

The long-standing anomaly of the muon $g-2$ could be due to
systematics in previous measurements or signal a departure from the SM
caused by new physics in loop contributions. Most likely (naive
scaling - NS), such new physics will manifest in the electron sector
with a $(a_e-a_e^{SM})/(a_\mu-a_\mu^{SM}) \simeq (m_e/m_\mu)^2$
suppression due to the different lepton masses.  A major experimental
effort is ongoing to clarify this issue and we expect new data in the
muon sector to be available in a few years. In this paper, we have
shown that - on a similar timescale - the experimental measurement of
$a_e$ can provide key information since the precision that is
attainable is comparable with NS expectations. From the
experimental point of view, the most critical challenge is a sub-ppb
determination of the $h/M$ quotient.  Atom interferometry can provide
this measurement and, through the Rydberg relationship, measure the
fine structure constant independently of $a_e$. Recent advances in
metrology and, in particular, the revised measurement of the electron
mass, reduced the systematics due to the $m_e/m_u$ ratio to a level
appropriate for this goal. Among atom candidates, $^{4}\mathrm{He}$
is particularly appealing due to the outstanding accuracy (0.015~ppb)
obtained on $M_{He}/m_u$.

\section*{Acknowledgments}
It is a great pleasure to thank M.~Passera for bringing
Ref.~\cite{Giudice:2012ms} to our attention and for many useful
remarks in the early stage of this work. We are indebted to
P.~Paradisi and M.~Passera for careful reading of the manuscript. We
wish to express our gratitude to K.~Blaum, P.~Clad\'e, H.~M\"uller and
W.~Vassen for useful information and discussions.


\end{document}